\documentclass[10pt]{article}

\usepackage{graphicx}%
\usepackage{multirow}%
\usepackage{amsmath,amssymb,amsfonts}%
\usepackage{amsthm}%
\usepackage{mathrsfs}%
\usepackage[title]{appendix}%
\usepackage{xcolor}%
\usepackage{textcomp}%
\usepackage{manyfoot}%
\usepackage{booktabs}%
\usepackage{algorithm}%
\usepackage{algorithmicx}%
\usepackage{algpseudocode}%
\usepackage{listings}%

\theoremstyle{thmstyleone}%
\theoremstyle{thmstyletwo}%

\theoremstyle{thmstylethree}%

\usepackage[all]{nowidow}
\usepackage{pgfplots}
\usepackage{pgfplotstable}
\usetikzlibrary{pgfplots.groupplots,arrows,automata,positioning,arrows.meta,patterns}
\usepgfplotslibrary{dateplot,fillbetween,statistics}
\pgfplotsset{compat=1.18}

\definecolor{magenta}{RGB}{247,99,186} 
\definecolor{purple}{RGB}{106,50,159} 
\definecolor{darkblue}{RGB}{4,64,118} 
\definecolor{lightblue}{RGB}{28,158,255} 
\definecolor{green}{RGB}{73,153,38}
\definecolor{pink}{RGB}{241,147,147}
\definecolor{lightgreen}{RGB}{132,196,104}
\definecolor{darkpink}{RGB}{244,115,115}
\definecolor{pastelgreen}{RGB}{134,191,145}
\definecolor{pastelorange}{RGB}{241,173,102}

\tikzstyle{count_aff} = [
smooth,
purple,
thick, mark=none, 
draw opacity=0.8
]
\tikzstyle{fitt_aff1} = [
smooth,
magenta,
thick, mark=none, 
dashed, 
draw opacity=0.8
]
\tikzstyle{fitt_aff2} = [
smooth,
magenta,
thick, mark=none, 
dashed, 
draw opacity=0.8
]
\tikzstyle{node_aff} = [
circle,inner sep=1pt,
minimum size=0.5pt, draw=black, thick,
fill=magenta!50,
]

\tikzstyle{count_aff_diff} = [
smooth,
darkblue,
thick, mark=none, 
draw opacity=0.8
]
\tikzstyle{fitt_aff_diff1} = [
smooth,
lightblue,
thick, mark=none, 
dashed, 
draw opacity=0.8
]
\tikzstyle{fitt_aff_diff2} = [
smooth,
lightblue,
thick, mark=none, 
dashed, 
draw opacity=0.8
]
\tikzstyle{node_aff_diff} = [
circle,inner sep=1pt,
minimum size=0.5pt, draw=black, thick,
fill=lightblue!50,
]

\tikzstyle{count_collab} = [
smooth,
darkgreen,
thick, mark=none, 
draw opacity=0.8
]
\tikzstyle{fitt_collab1} = [
smooth,
lightgreen,
thick, mark=none, 
dashed, 
draw opacity=0.8
]
\tikzstyle{fitt_collab2} = [
smooth,
lightgreen,
thick, mark=none, 
dashed, 
draw opacity=0.8
]
\tikzstyle{node_collab} = [
circle,inner sep=1pt,
minimum size=0.5pt, draw=black, thick,
fill=lightgreen!50,
]

\tikzstyle{count_collab_diff} = [
smooth,
darkpink,
thick, mark=none, 
draw opacity=0.8
]
\tikzstyle{fitt_collab_diff1} = [
smooth,
pink,
thick, mark=none, 
dashed, 
draw opacity=0.8
]
\tikzstyle{fitt_collab_diff2} = [
smooth,
pink,
thick, mark=none, 
dashed, 
draw opacity=0.8
]
\tikzstyle{node_collab_diff} = [
circle,inner sep=1pt,
minimum size=0.5pt, draw=black, thick,
fill=pink!50,
]

\tikzstyle{distr_index} = [
smooth,
green,
thick, mark=none, 
draw opacity=0.8
]

\tikzstyle{distr_index_err} = [
fill=green!50,
fill opacity=0.5,
]

\tikzstyle{distr_index2} = [
smooth,
orange,
thick, mark=none, 
draw opacity=0.8
]
\tikzstyle{distr_index_err2} = [
fill=orange!50,
fill opacity=0.5,
]

\tikzstyle{distr_index3} = [
smooth,
red,
thick, mark=none, 
draw opacity=0.8
]
\tikzstyle{distr_index_err3} = [
fill=red!50,
fill opacity=0.5,
]

\tikzstyle{distr_index4} = [
smooth,
blue,
thick, mark=none, 
draw opacity=0.8
]

\usepackage[eulergreek]{sansmath}
\pgfplotsset{
  tick label style = {font=\sansmath\sffamily},
  every axis label = {font=\sansmath\sffamily},
  legend style = {font=\sansmath\sffamily},
  label style = {font=\sansmath\sffamily},
  title style = {font=\sansmath\sffamily}
}
\tikzset{every picture/.style={/utils/exec={\sffamily}}}

\pgfplotsset{
    restrict x to domain**/.code args={#1:#2}{
        \pgfkeysalso{/pgfplots/x coord trafo=#1}
        \let\numericxmin\pgfmathresult
        \pgfkeysalso{/pgfplots/x coord trafo=#2}
        \let\numericxmax\pgfmathresult
        \pgfkeysalso{/pgfplots/restrict x to domain={\numericxmin}:{\numericxmax}}
    }
}

\usepackage{mathtools,multirow,multicol,graphicx,comment,standalone,array,amsmath,amssymb,blkarray,colortbl,xspace,booktabs}
\usepackage{subcaption}
\newcolumntype{C}{>{\raggedleft\arraybackslash}p{3mm}}
\usepackage{tikz,pgfplots,pgfplotstable}
\pgfplotsset{compat=1.16,set layers, mark layer=axis tick labels}

\definecolor{NDred}{HTML}{AA272F}

\usepackage{pgfplots}
\usepackage{pgfplotstable}
\usetikzlibrary{pgfplots.groupplots,arrows,automata,positioning,arrows.meta,patterns}
\usepgfplotslibrary{dateplot,fillbetween,statistics}
\pgfplotsset{compat=1.18}

\tikzstyle{node} = [
circle,inner sep=1pt,
minimum size=3pt, draw=orange, thick,
fill=orange!50,
]
\usetikzlibrary{shapes,snakes}

\definecolor{magenta}{RGB}{247,99,186} 
\definecolor{purple}{RGB}{106,50,159} 
\definecolor{darkblue}{RGB}{4,64,118} 
\definecolor{lightblue}{RGB}{28,158,255} 
\definecolor{green}{RGB}{73,153,38}
\definecolor{pink}{RGB}{241,147,147}
\definecolor{lightgreen}{RGB}{132,196,104}
\definecolor{darkgreen}{RGB}{31,86,7}
\definecolor{darkpink}{RGB}{244,115,115}
\definecolor{lightpink}{RGB}{225,145,192}
\definecolor{lightblue}{RGB}{152,198,239}
\definecolor{lightgreen2}{RGB}{176,225,66}
\definecolor{bordeaux}{RGB}{95,2,31}
\definecolor{magenta}{RGB}{195,28,80}
\definecolor{ciano}{RGB}{36,179,239}
\definecolor{purple}{RGB}{103,78,167}
\definecolor{inter}{RGB}{91,132,177}
\definecolor{intra}{RGB}{252, 118, 106}
\definecolor{interdark}{RGB}{58,88,120}
\definecolor{intradark}{RGB}{203,82,71}
\definecolor{IF1}{RGB}{117, 121, 71}
\definecolor{IF2}{RGB}{108, 145, 126}
\definecolor{IF3}{RGB}{39, 89, 96}
\definecolor{IF4}{RGB}{108, 145, 126}
\definecolor{lightIF3}{RGB}{167,192,194}

\usepackage{booktabs}       
\usepackage{nicematrix}
\usepackage{booktabs}
\usepackage{caption}

\newcommand{\methods}{\hyperref[sec:methods]{Methods}\xspace}

\usetikzlibrary{shapes}

\usepackage{authblk}
\usepackage{url}
\newcommand{\bmhead}[1]{\paragraph{\textbf{#1}}{}}
\usepackage{xspace}
\usepackage{hyperref}
\newcommand{\RNum}[1]{\MakeUppercase{\romannumeral #1}}
\usepackage[margin=1in]{geometry}

\title{\large The Effects of Remote Working on Scientific Collaboration and Impact}

\author[1,2]{Sara Venturini}
\author[3]{Satyaki Sikdar}
\author[1]{Martina Mazzarello}
\author[4]{Francesco Rinaldi}
\author[5,6]{Francesco Tudisco}
\author[1,7]{Paolo Santi}
\author[8]{Santo Fortunato}
\author[1,9]{Carlo Ratti}

\affil[1]{Senseable City Laboratory, Massachusetts Institute of Technology, Cambridge, MA, USA}
\affil[2]{Network Science Institute, Northeastern University, Boston, MA, USA}
\affil[3]{Department of Computer Science, Loyola University Chicago, Chicago, IL, USA}
\affil[4]{Department of Mathematics “Tullio Levi-Civita”, University of Padova, Padova, Italy}
\affil[5]{School of Mathematics, The University of Edinburgh, Edinburgh, Scotland, UK}
\affil[6]{School of Mathematics, Gran Sasso Science Institute, L’Aquila, Italy}
\affil[7]{Instituto di Informatica e Telematica del CNR, Pisa, Italy}
\affil[8]{Luddy School of Informatics, Computing, and Engineering, Indiana University, Bloomington, IN, USA}
\affil[9]{Department ABC, Politecnico di Milano, Milan, Italy}

\date{}

\begin{document}

\maketitle

\begin{abstract}
The COVID-19 pandemic shifted academic collaboration from in-person to remote interactions. This study explores, for the first time, the effects on scientific collaborations and impact of such a shift, comparing research output before, during, and after the pandemic. Using large-scale bibliometric data, we track the evolution of collaboration networks and the resulting impact of research over time. Our findings are twofold: first, the geographic distribution of collaborations significantly shifted, with a notable increase in cross-border partnerships after 2020, indicating a reduction in the constraints of geographic proximity. Second, despite the expansion of collaboration networks, there was a concerning decline in citation impact, suggesting that the absence of spontaneous in-person interactions—which traditionally foster deep discussions and idea exchange—negatively affected research quality. As hybrid work models in academia gain traction, this study highlights the need for universities and research organizations to carefully consider the balance between remote and in-person engagement.
\end{abstract}

\section{Introduction}\label{sec:introduction}

The global COVID-19 pandemic, besides its profound impact on public health, transformed academic and professional life, affecting the way science is conducted~\cite{myers20,maher2021covid,sikdar2024we,fry2020consolidation,viglione2020women}. It has been widely regarded as a natural experiment in new ways of working, particularly in how it disrupted co-location in physical spaces. One key variable, face-to-face collaboration, was abruptly removed during widespread lockdowns and replaced with remote forms of interaction. As the COVID-19 pandemic emergency ended, in-person exchanges gradually resumed, but the attitude towards increased use of virtual meeting and remote collaboration tools persisted. These changes are likely to have enduring consequences on collaboration networks and scientific production for years to come~\cite{maher2021covid}. 
 
On the one hand, the pandemic prompted an unprecedented pivot in research agendas. Scholars across disciplines, not only in epidemiology, redirected their focus to COVID-19-related questions, driven by an urgent need to inform policy and accelerate solutions \cite{sikdar2024we}. The crisis also acted as a catalyst of scientific novelty by stimulating original and multidisciplinary work \cite{liu2022pandemics}. In fact, Cunningham et al. \cite{cunningham2021collaboration} highlight that the multidisciplinary nature of SARS-CoV-2 research rapidly reshaped collaboration patterns. 

On the other hand, the absence of face-to-face encounters fundamentally altered the mechanisms through which ideas are generated and disseminated \cite{olson2000distance}. Prior research shows that while remote communication may sustain productivity, it tends to reduce creativity and the emergence of disruptive ideas \cite{brucks2022virtual, lin2023remote,gorlich2023creativity}.
The decline in “weak ties” during the lockdowns—connections that are crucial for transmitting novel information—has been documented across both academic and corporate environments \cite{carmody2022effect,yang2022effects}. Studies further confirm that strong, co-located ties tend to dominate remote collaboration, often at the expense of serendipitous interactions and cross-disciplinary innovation \cite{zeng2021fresh}.

Beyond individual collaborations, the pandemic also reshaped the broader spatial and institutional landscape of science. Research has long shown that the intensity of collaboration between cities decreases with geographic distance, following gravity-like laws \cite{pan2012world}. Recent work points to an inversion of pre-pandemic trends: while new collaborations had been in decline, the pandemic years saw growth in cross-institutional and international ties~\cite{fu2023collaboration, cunningham2021collaboration}. At the same time, the impact of these collaborations, measured in citations, after an initial burst driven mainly by COVID-19-related research, seems to show a decline in the early post-pandemic phase~\cite{maillard22, uddin23, kim24}.

While these studies provide important insights, they have largely focused on the immediate aftermath of the pandemic. Much less is known about the longer-term effects of these structural changes on research productivity and scholarly impact. Our study addresses this gap by providing the first large-scale, longitudinal analysis of how collaboration networks and academic output evolved across three phases: before, during, and after the pandemic. Using bibliometric data from OpenAlex, along with preprints from the ArXiv repository (1.6 million preprints and 1.2 million authors), we examine shifts in collaboration geography, institutional ties, and the citation impact of scholarly work. Our analysis shows that the geographic distribution of collaborations shifted significantly, with a notable increase in cross-border partnerships after 2020. On the other hand, this expanded global footprint of collaborations seems to be accompanied by a decline in citation impact, hinting to a possible negative effect of remote collaboration on the quality of the research outcome. 

In sum, the COVID-19 pandemic simultaneously stimulated new lines of research, expanded cross-institutional collaborations, and undermined some of the creative mechanisms traditionally fueled by in-person scientific exchange. As the world moves into a hybrid era of work, understanding the long-term implications of these shifts—balancing flexibility with the irreplaceable value of physical co-location—remains a critical challenge for universities, funding bodies, and research organizations \cite{john2022enhancing,charpignon2023navigating}.

\section{Results}\label{sec:results}

We utilize the bibliometric dataset \texttt{OpenAlex} \cite{priem2022openalex} and restrict our analysis to preprints published in the \texttt{ArXiv} repository, to ensure more reliable publication dates, as journal publication dates would introduce additional hurdles in estimating the period when the collaboration took place. See \methods for details.\\
To investigate the impact of the COVID-19 pandemic and remote work on team collaborations, we examine the evolution of the collaboration network between authors, emphasizing their geographical proximity. The collaboration network is constructed as a graph where nodes represent authors and edges indicate co-authorship on preprints. We develop a model to explain the observed collaboration patterns and examine how these changes relate to the citation impact of the resulting work. All the measures used in these sections are formally defined in \methods.

\begin{figure}
\centering
\includegraphics[scale=0.8]{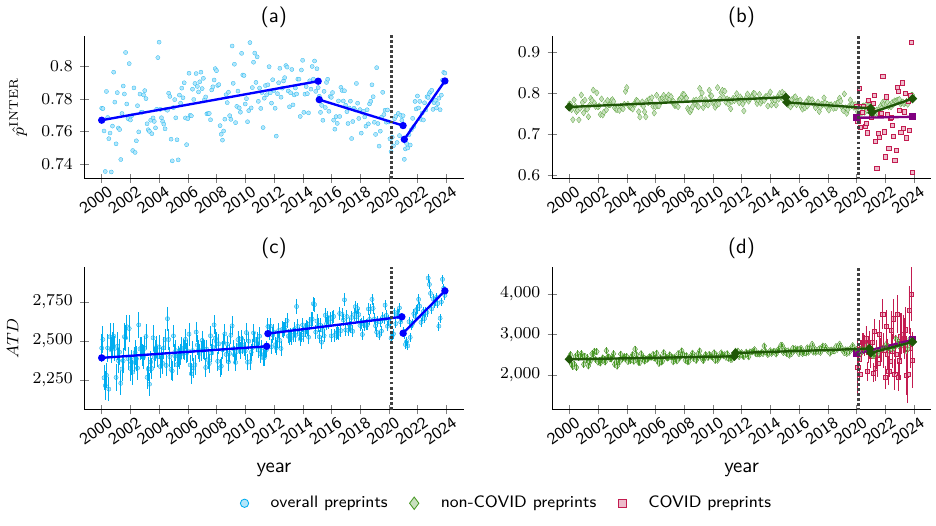}
\caption{\textbf{Monthly trends in inter-institution collaboration and geographic team distance.} Monthly average of the fraction of inter-institution collaborations for:
\textbf{(a)} all preprints, and
\textbf{(b)} COVID-19-related and non-COVID-19-related preprints (in magenta and green, respectively).
The fraction of intra-institution collaboration shows the opposite trend.
\textbf{(c)} and \textbf{(d)} show the monthly average team geographic distance (in kilometers) with standard error bars for the same groups.
Vertical dotted lines mark the onset of the COVID-19 pandemic (March 2020).
A piecewise polynomial regression with two breakpoints is fitted for all preprints and non-COVID-19-related preprints, and a single polynomial model without breakpoints is fitted for COVID-19-related preprints, minimizing squared error.}
\label{fig:panel0}
\end{figure}

In this work, we investigate the evolution over time of the collaboration network between scholars. 
Focusing on the geographical proximity of collaborations, we define a collaboration between two coauthors as intra-institution if both coauthors are affiliated with the same institution, and inter-institution otherwise (see \methods\ for details). Since preprints may involve more than two authors, we represent each preprint as a fully connected clique: an undirected edge is created between every pair of coauthors. An edge exists between two authors if they have collaborated at least once, and its weight reflects the number of such collaborations.
In Fig.~\ref{fig:panel0}a, we illustrate how the fraction of inter-institutional collaborations has evolved over time.
A piecewise polynomial model with two breakpoints is fitted to the data by minimizing the squared error. 
We observe an increasing trend from $2000$ to $2015$, followed by a decreasing trend from $2015$ to $2021$, and a subsequent marked increase thereafter. Intra-institution collaborations exhibit the opposite trend. This pattern suggests that the COVID-19 pandemic encouraged scholars to collaborate more frequently with researchers from different institutions.
In Fig.~\ref{fig:panel0}b, we report the same analysis, distinguishing between non-COVID-19 and COVID-19-related preprints. Since COVID-19-related preprints represent only a small fraction of the total, the overall trend remains largely unchanged.\\
However, the \textit{inter-institution} analysis may not provide sufficient granularity, as it encompasses varying distances between institutions. 
To capture this variation more precisely, in Fig.~\ref{fig:panel0}c, we plot the monthly mean of the average team distance ($ATD$) of the preprints, which reflects the average geographical distance between the institutions of coauthors (see \methods\ for details). A piecewise polynomial model with two breakpoints is fitted to the data by minimizing the squared error. We observe increasing trends from $2000$ to $2011$, and from $2011$ to $2021$, with a sharp rise thereafter. This pattern suggests that the COVID-19 pandemic prompted scholars to collaborate more frequently with geographically distant colleagues.
To evaluate the role of COVID-19-related papers in shaping this behavior, Fig.~\ref{fig:panel0}d presents the same analysis applied separately to non-COVID-19 and COVID-19-related preprints. The overall trend remains largely consistent, indicating that the observed increase in $ATD$ is not solely driven by COVID-19-related research.\\
\begin{figure}
\centering
\includegraphics[scale=0.8]{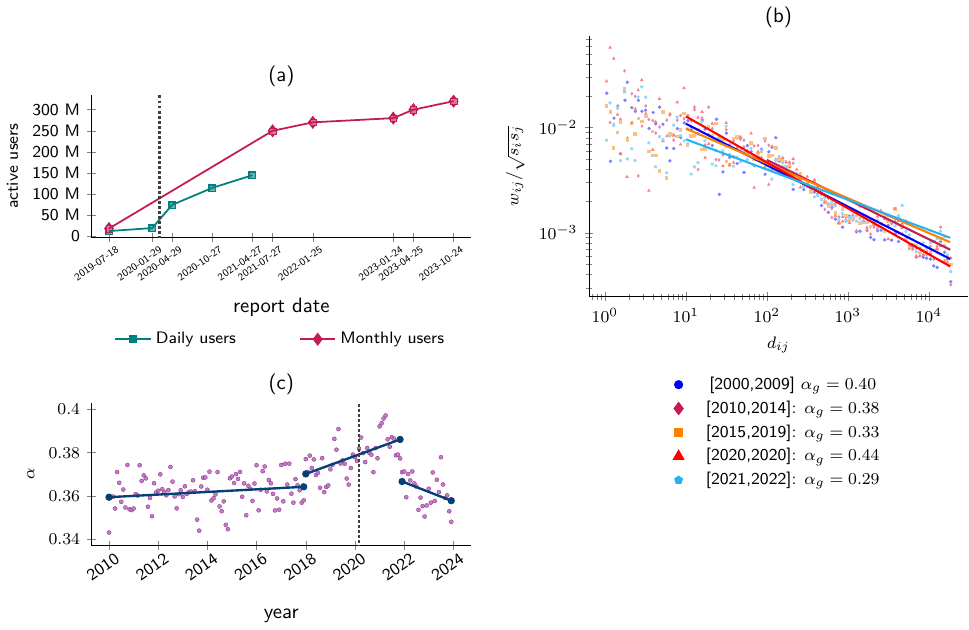}
\caption{
\textbf{Temporal patterns in Microsoft Teams usage and distance-dependent collaboration.}
\textbf{(a)} Daily (green) and monthly (red) active users on Microsoft Teams from July 2019 to October 2023, based on public earnings reports. The onset of the COVID-19 pandemic (March 2020) is indicated by a vertical dotted line. 
\textbf{(b)} Collaboration strength between institutions, normalized by institution productivity, decreases with geographic distance and follows a gravity law. Colors correspond to different time periods, each fitted with a power-law decay; $\alpha_g$ denotes the distance decay exponent.  
\textbf{(c)} Temporal evolution of the optimized geographical proximity parameter $\alpha$ in the collaboration network, according to our model. A piecewise polynomial with two breakpoints is fitted by minimizing the squared error. The vertical dotted line marks the onset of the COVID-19 pandemic (March 2020).
}
\label{fig:panel1}
\end{figure}
To compare our analysis of geographical proximity in collaborations with broader trends in remote work, Fig.~\ref{fig:panel1}a shows the number of daily and monthly active users on Microsoft Teams, based on data from Microsoft’s public earnings reports. We observe a sharp increase in the number of active users beginning in April $2020$, coinciding with the onset of the COVID-$19$ pandemic. By April $2021$, the number of users had nearly doubled compared to the previous year. Notably, the upward trend continues even after the acute phase of the pandemic, indicating a sustained shift toward remote work. These findings highlight how COVID-$19$ has had a lasting impact on work and collaboration practices.\\
Building on this, we model the evolution of researchers' collaboration patterns, with particular attention to the role of institutional affiliation and the geographic distance between coauthors (see \methods).
In particular, we would like to reproduce two quantities: the probability of intra-institution collaborations (${p}^{INTRA}$) and the $ATD$. 
Our approach is inspired by the work of \cite{pan2012world}, which demonstrated that collaboration strength between cities decreases with geographic distance and follows gravity laws. As shown in Fig.~\ref{fig:panel1}b, we confirm a similar decay pattern in our dataset, where the normalized collaboration strength between institutions is inversely proportional to distance, consistent with a gravity model. 
The key differences in our approach are the use of institutions instead of cities and the incorporation of the square root of the product of the collaboration strengths.
As shown in Fig.~A1, the self-loop weight increases approximately linearly with institutional strength, which supports using institutional strengths for intra-institution collaborations. Therefore, applying the square root of the product of strengths for inter-institution pairs provides a natural and consistent extension.
In the main text, we focus on the simplest case of two-author papers (more than $30\%$ of the whole dataset), while in the Supplementary Information (Fig.~A2) we test the model's robustness by extending the analysis to the full preprint dataset, treating each paper as a clique; and considering papers as hyperedges, but limit experiments to three-author papers (more than $26\%$ of the whole dataset) due to computational constraints.
We initialize the institutional network $H_0$ using the preprints published in the $10$-year window between $2000$ and $2009$, as described in \methods.
The institutional network is a graph where nodes represent research institutions and edges indicate collaborations between them, defined by co-authorship of at least one preprint by affiliated authors.
Therefore, we fix the set of institutions $I$ which have published at least one preprint in the window, and we calculate their distances and their initial strengths. Specifically, we are considering a set of $3.4$K institutions.
The model operates as follows: for any given pair of authors, there are two possibilities — either they belong to the same institution or to two different institutions. In the first case, this corresponds to a self-edge in the institutional network, with a probability proportional to the institution's strength, consistent with observed data (see Supplementary Fig~S8).
In the second case, the collaboration forms a proper edge between two different institutions. The probability of this edge forming depends on the square root of the product of the two institutions’ strengths (inspired by the preferential attachment model \cite{barabasi1999emergence}), and is inversely proportional to a power of the geographical distance between them.\\
For each month $t>0$, we use the total number of published papers $W_t$ in the dataset, and we let the initial network grow with these probabilities 
\begin{equation}
    \label{probs}
    \begin{cases}
    P^{ij}(\alpha,\beta,\gamma) = \displaystyle \frac{B_{ij}}{N(\alpha,\beta,\gamma)} \quad \forall i,j \in I \text{ s.t. } i<j,\\
    P^{i}(\alpha,\beta,\gamma) = \displaystyle \frac{B_{i}}{N(\alpha,\beta,\gamma)} \quad \forall i \in I,
    \end{cases}
\end{equation}
where  
\begin{equation}
    \begin{cases}
    B^{ij}(\alpha,\beta) = \beta  \displaystyle \frac{\sqrt{s_is_j}}{(d_{ij}+100)^{\alpha}} \quad \forall i,j \in I \text{ s.t. } i<j,\\
    B^{i}(\gamma) = \gamma s_i \quad \forall i \in I,
    \end{cases}
\end{equation}
and we normalize them as follows 
\begin{equation}
N(\alpha,\beta,\gamma) = \sum_{i,j \in I: i<j} B^{ij}(\alpha,\beta) + \sum_{i \in I} B^{i}(\alpha,\beta,\gamma).
\end{equation}
$\alpha, \beta, \gamma$ are variables that we constrain to be non-negative real numbers. 
See \methods for the extension of these probabilities to the hyperedge case.
We emphasize that, due to the normalization condition, the model has only two independent variables. However, reducing the number of variables further would eliminate box constraints, increasing the complexity of the solution. Furthermore, after optimizing the model with all variables, we observed that $\beta$ and $\gamma$ are approximately proportional to each other (see Supplementary Fig.~A3). 
Therefore, we optimize the problem by focusing only on the exponent $\alpha$, which makes also the results more interpretable. 
For each month $t>0$, we aim to fit the model to the ${p}^{INTRA}$ and the $ATD$ from the data. To achieve this, we optimize the sum of their relative errors per month. Since the institutional network evolves over time, the institutions' strengths are also updated on a monthly basis.
The strength of an institution is defined as the sum of the weights of all edges connected to it, reflecting its collaborative productivity (see \methods for a formal definition).
In Fig.~\ref{fig:panel1}c, we show the evolution of the parameter $\alpha$ over time to assess the impact of the COVID-$19$ pandemic on the role of institutional distance in shaping research collaborations. 
Lower values of $\alpha$ indicate that distance plays a smaller role in collaboration formation, and vice versa. A piecewise polynomial model with two breakpoints is fitted to the data by minimizing the squared error.
The results show that the importance of distance in collaborations remained relatively stable until around $2018$, after which it increased sharply up to $2021$. Following $2021$, however, it began to decline and continued to do so through the end of $2023$.
These findings are consistent with the patterns observed above, reinforcing the idea that the pandemic reduced the influence of geographical distance in academic collaborations.
In order to further validate the proposed model, we perform several simulations starting from the initial institutions' network $H_0$, growing it according to the probabilities in \eqref{probs}.
In Supplementary Fig.~A4, respectively, we show the average quantities obtained across $10$ runs, compared with the monthly quantities from the data (see Figs.~S5–S6 for robustness tests on the full dataset and three-author papers).\\  
\begin{figure}
\centering
\includegraphics[scale=0.8]{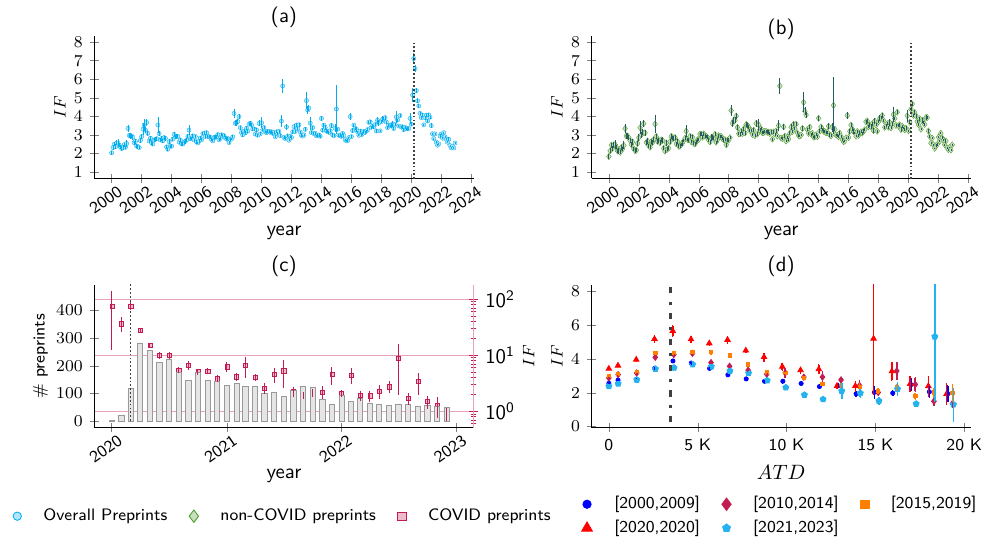}
\caption{
\textbf{One-year impact factor and its relationship with team distance in preprints.}
\textbf{(a)} Monthly average number of citations ($IF$) received by preprints within one year from publication, with standard error bars.  
\textbf{(b)} Same as (a), restricted to non-COVID-19-related preprints.  
\textbf{(c)} Monthly counts of COVID-19-related preprints (gray bars) and their average one-year citation counts (magenta, logarithmic scale) with standard error bars. 
Vertical dotted lines indicate the onset of the COVID-19 pandemic (March 2020).
\textbf{(d)} Relationship between average team distance ($ATD$) and $IF$ for preprints, split into five consecutive time periods. 
The loosely dash-dotted vertical line marks the $ATD$ of the authors of this study ($3474.61$ km).
}
\label{fig:panel2}
\end{figure}
Finally, we investigate whether changes in collaboration patterns influence the works produced by these teams. In Fig.~\ref{fig:panel2}a, we report the one-year Impact Factor ($IF$)—defined as the number of citations received within one year of publication—for preprints. A pronounced peak appears in January $2020$, which disappears when excluding preprints tagged as COVID-19 related. This behavior is explained by Fig.~\ref{fig:panel2}c, which shows the explosive growth in the production of COVID-19-related papers starting in March 2020. These papers frequently cite the earliest coronavirus-related preprints published in the first two months of 2020. This is in line with the \textit{first-mover advantage}, where the early explorers of a field benefit from greater prominence~\cite{newman09}. 
In contrast, the $IF$ trend for non-COVID-19-related preprints (also shown in Fig.~\ref{fig:panel2}b) exhibits a steady increase from $2000$ to $2008$, followed by a slower increase between $2008$ and $2020$, and a decline thereafter—without any sharp peak.
In Supplementary Fig.~A7, we present the same analysis, this time considering the number of citations received within two years of publication for preprints, which shows a similar trend.\\
In Fig.~\ref{fig:panel2}d, we report the relationship between $ATD$ and $IF$ using a scatter plot of average values, grouped into five consecutive time periods (see Supplementary Tab.~S1). The plot suggests that there is a preferred range of $ATD$ associated with higher $IF$ values. Notably, the COVID-19 period ($[2020,2020]$, shown in red) exhibits generally higher $IF$ values compared to earlier periods.

\section{Discussion}\label{sec:discussion}

Our study provides evidence that the COVID-19 pandemic reshaped the geography and impact of scientific collaboration. By analyzing papers from a large-scale bibliometric dataset before, during, and after the pandemic, we find that remote work and virtualization broadened the spatial reach of collaborations, while reducing their scientific impact. This apparent paradox—greater diversity in authorship combined with lower citation outcomes—highlights the dual role of digital tools: they can expand networks, but they cannot fully substitute for the creativity and serendipity fostered by in-person encounters.
These findings extend and nuance prior work. Earlier studies suggested that remote communication can sustain productivity but often at the expense of disruptive and original contributions \cite{brucks2022virtual,lin2023remote,gorlich2023creativity}. Our results confirm this pattern at scale, showing that the pandemic accelerated the erosion of “weak ties” \cite{carmody2022effect,yang2022effects}, which are essential for novel idea generation. While previous research has documented the gravity-like laws of scientific collaboration \cite{pan2012world}, we demonstrate that the pandemic effectively altered these constraints, producing an unprecedented rise in long-distance and cross-institutional partnerships \cite{cunningham2021collaboration,fu2023collaboration}. Yet, the expected gains in impact from greater diversity \cite{zeng2021fresh} did not materialize, pointing to the unique value of face-to-face interactions in sustaining high-quality science.

This study is not without limitations. Our analysis is based on bibliometric indicators, which capture publication and citation patterns but not the full richness of the research process, including informal mentorship, idea incubation, or experimental collaboration. Furthermore, the post-pandemic period analyzed here is still relatively short; longer-term monitoring will be essential to determine whether the decline in impact persists or stabilizes as hybrid practices mature. Future work could complement bibliometric analysis with surveys, ethnographic studies, or experimental interventions to better capture the social mechanisms driving collaboration.

The implications of the results presented in this study extend beyond academia. For funding bodies and research organizations, the findings call for careful consideration of hybrid work models that combine the inclusivity and reach of digital platforms with mechanisms to recreate the depth and creativity of in-person exchanges. For policymakers, our evidence underscores that geographic proximity still matters, not only for convenience but also for the generative processes that underpin impactful scientific discovery. Similar dynamics may also apply to innovation in industry and R\&D, where distributed teams face analogous challenges.

Ultimately, our results suggest that while the sudden shift to remote collaboration tools caused by the COVID-19 pandemic democratized access to scientific networks, it simultaneously weakened some of the creative mechanisms that thrive in physical proximity. The challenge for the next decade is to design new organizational models that preserve the benefits of global reach while reinvigorating the serendipity of in-person encounters—a balance that will be crucial for the future of scientific progress.

\section*{Methods} \label{sec:methods}


\bmhead{Dataset} \label{sec:methods-data-datasets}
We analyze works from the December $2024$ snapshot of the bibliometric dataset \texttt{OpenAlex} \cite{priem2022openalex}.
We restrict our analysis to preprints published on the \texttt{ArXiv} repository, between $2000$ and $2023$. We consider $1.6$M preprints and $1.2$M authors. In Fig.~A8 we show the time evolution of the number of preprints and authors.
We link the papers in the \texttt{ArXiv} repository with the ones in the \texttt{OpenAlex} dataset matching their DOIs, locations, and open-access data. We discard matchings where the papers have a different number of authors and title similarity lower than $95\%$.
We use the \texttt{ArXiv} repository to rely on the preprints' publication dates and for the divisions into categories. 
From the \texttt{OpenAlex} dataset, we use the list of disambiguated authors and their affiliations (ID, country, location).
The distribution of preprints into categories is in Tab.~S2 and Fig.~A9. 
Since we focus on teams, we restrict the analysis to preprints with at least two authors, discarding single-author papers, and with at most $30$ authors. 
We further filter out preprints with missing concepts or references.
For citation-based calculations, we use citation data from the entire \texttt{OpenAlex} dataset ($2000$–$2024$). 

\bmhead{Authors' affiliations} \label{sec:methods-data-affiliations}
Each author has a dynamic list of affiliations that changes based on paper authorship.
For each author, we analyze their list of published works in the whole \texttt{OpenAlex} and assign the most frequent affiliation for each year, picking randomly if multiple affiliations have the same frequency. This prevents the authors from having multiple affiliations. Also, it results in a chronological list of universities, allowing us to identify the author's change in affiliation. Therefore, we assign to each preprint's authors their affiliation in the correspondent period. 
We retain only the preprints that include complete information about all authors' institutions.\\
Furthermore, since the institutions in \texttt{OpenAlex} have a hierarchical structure and to avoid treating two universities as different when they are the same, we substitute each university with its parent in the hierarchy, and with a random one if they have multiple parents. 


\bmhead{COVID-related preprints identification} \label{sec:methods-data-categories}
We identify works related to the COVID-19 pandemic by examining \texttt{OpenAlex} concepts and their titles. A preprint is considered relevant to COVID-19 if it is tagged with the concepts \texttt{Coronavirus disease 2019} (\texttt{COVID-19})(\texttt{C3008058167)}, \texttt{2019-20 coronavirus outbreak} (\texttt{C3006700255}), or \texttt{Severe acute respiratory syndrome coronavirus 2 (SARS-CoV-2)}\\(\texttt{C3007834351}), or if its title contains the word \texttt{covid}. This results in $4.8$K COVID preprints. The number of COVID-19 and non-COVID-19 preprints is shown in Fig.~A8c.



\bmhead{Distances between institutions} \label{sec:methods-dist}
We calculate the great-circle distance between two institutions using their latitude and longitude coordinates, which gives the shortest path over the Earth's surface — approximated as a sphere. We identify multiple institutions as the same if they are closer than $1$ km.  This way we end up with  $20.9$K institutions. 
Following the approach in \cite{pan2012world}, we project the collaboration network onto the network of institutions, in that each work represents a weighted clique connecting the unique institutions of the authors.
For instance, a preprint between authors of the same institution will be a self-loop of weight one, and a preprint between $k$ authors of $k$ different institutions will be a clique between them with all $\binom{k}{2}$ edges of weight $\frac{1}{\binom{k}{2}}$.

\bmhead{Average team distance} \label{sec:methods-dist}
For each preprint, we calculate the Average Team Distance (ATD), which is the average pairwise distance of the institutions of the authors:
\begin{equation}
\label{eq:ATD}
ATD(\omega) = \displaystyle \frac{\sum_{(i,j) \in\ E(\omega)} d_{ij}}{|E(\omega)|},
\end{equation}
where $\omega$ is a preprint, $E(\omega)$ is the set of edges of the correspondent clique in the scholars' collaboration network, $d_{ij}$ is the distance between the institutions of authors $i$ and $j$.\\


\bmhead{Model} \label{sec:methods-model}
We model the evolution of researchers' collaborations, focusing on the influence of their institutions and the geographic distance between them. In particular, we aim at reproducing the following two quantities:
\begin{itemize}
    \label{props}
    \item Probability of intra-institution collaborations in each month $t$, calculated as the fraction of the weights of intra-institution collaborations, divided by the total weight of the collaboration graph:
    \begin{equation}
    \label{eq:Pintra}
        \hat{p}^{INTRA}_{t} =  \displaystyle \frac{W^{INTRA}_{t}}{W_{t}}.
    \end{equation}
    Here $W_{t}$ is the total weight of the collaboration network in month $t$, and\\ $W^{INTRA}_{t} = \sum_{(i,j) \in E_{t} : i,j \text{ in the same institution}} w_{ij}$, with $E_{t}$ representing the set of edges in the collaboration network in month $t$. Consequently, the probability of inter-institution collaboration is $\hat{p}^{\mathrm{INTER}}_{t} = 1 - \hat{p}^{\mathrm{INTRA}}_{t}$.
    \item Monthly mean of the average team distance $\hat{ATD}_{t}$ in each month $t$. For each preprint, we calculate its ATD as described above, and for each month, we compute the mean across all preprints published in that month.
\end{itemize}
As in the main text, we focus on two-author papers here, but extensive results, including those extended to the entire preprint dataset and considering papers as hyperedges, are provided in the Supplementary Information.\\
We initialize the network of institutions $H_0$ using the preprints published in the $10$-year window between $2000$ and $2009$. Therefore, we fix the set of institutions $I$, as the one comprising institutions that have published at least one preprint in the window, and we calculate their distances, and their initial strengths as follows
\begin{equation}
    s^i_0 = \sum_{e : i \text{ extreme of } e} w^e  \quad \forall i \in I,
\end{equation}
where $w^e$ is the weight of edge $e$, and the sum runs over all the edges $e$ having $i$ as extreme.
For each month $t>0$, we use the total number of published papers in that month $W_t$ in the dataset, and we let the initial network grow following the probabilities in \eqref{probs}.
For each month $t>0$, we want to fit the model to the two data values in \eqref{eq:ATD}-\eqref{eq:Pintra}, therefore we need to calculate an explicit expression for them for the model. 
The expression for ${p}^{INTRA}$ corresponds to the sum of the probabilities of the self-loops
\begin{equation}
    p^{INTRA}_{t} (\alpha,\beta,\gamma) = \sum_{i \in I} P_{i} .
\end{equation}
The monthly mean of the ATD corresponds to the expected value
\begin{equation}
    ATD_{t} (\alpha,\beta,\gamma) = 
    \sum_{i,j \in I}d_{ij}P^{ij}_t + \sum_{i \in I}d_{ii}P^{i}_t = \sum_{i,j \in I}d_{ij}P^{ij}_t ,
\end{equation}
where the second equality follows from the
fact that $d_{ii}=0$.
We emphasize that the quantities are related to the collaboration graphs, whereas in the model we are growing the corresponding  network of institutions.
We fit the model to the data, minimizing the sum of the squared relative errors of these two quantities:
\begin{equation}
\min_{\alpha,\beta,\gamma \geq 0} 
\left( \frac{\hat{p}^{INTRA}_{t}-p^{INTRA}_{t}(\alpha,\beta,\gamma)}{\hat{p}^{INTRA}_{t}} \right)^2 +
\left( \frac{\hat{ATD}_{t}-ATD_{t}(\alpha,\beta,\gamma)}{\hat{ATD}_{t}} \right)^2 \quad \forall t>0.
\end{equation}
We optimize this quantity for each month, and since the institution network grows over time, we also update the institutions' strengths every month
\begin{equation}
    s^i_t = s^i_{t-1} + W_t \left(\sum_{i,j \in I}P^{ij}_{t-1} + 2\sum_{i \in I} P^{i}_{t-1} \right) \quad \forall t>0.
\end{equation}
We perform the optimization using a sequential least squares programming algorithm, and for each month starting the algorithm from the optimal solution of the previous month. 
For the first month, we pick the best solution among the ones obtained starting from $9$ randomly generated points and the point $(\alpha,\beta,\gamma) = (2,1,1)$ according to the final objective function values. After optimizing the model with all variables, we observed  that $\beta$ and $\gamma$ are approximately proportional to each other. Therefore, we fixed their ratio at the mean value and optimized the model by focusing solely on the variable $\alpha$.\\

\bmhead{Model extension to hyperedges} \label{sec:methods-model-extension}
The proposed model can be extended to hyperedges, thereby modeling author collaborations not as network cliques, but as hyperedges. For computational reasons, in the Supplementary Information, we restrict the model to works with exactly three authors.
In this case, the distinction is that, whereas previously a preprint could involve either two authors from the same institution or from different institutions, we now have three possible configurations: (1) all authors from the same institution, (2) all authors from different institutions, and (3) two authors from the same institution with the third from a different one.
As in equation~\eqref{probs}, for each month $t > 0$, we use the total number of published papers $W_t$ in the dataset, and we allow the initial network to grow according to the following probabilities:
\begin{equation}
    \begin{cases}
    \bar P^{ijk}(\alpha,\beta,\gamma) = \displaystyle \frac{\bar B_{ijk}}{\bar N(\alpha,\beta,\gamma)} \quad \forall i,j,k \in I \text{ s.t. } i<j<k,\\
    \bar P^{ij}(\alpha,\beta,\gamma) = \displaystyle \frac{\bar B_{ij}}{\bar N(\alpha,\beta,\gamma)} \quad \forall i,j \in I \text{ s.t. } i<j,\\
    \bar P^{i}(\alpha,\beta,\gamma) = \displaystyle \frac{\bar B_{i}}{\bar N(\alpha,\beta,\gamma)} \quad \forall i \in I,
    \end{cases}
\end{equation}
where  
\begin{equation}
    \begin{cases}
    \bar B^{ijk}(\alpha,\beta) = \beta  \displaystyle \frac{\sqrt[3]{s_is_js_k}}{(d_{ijk}+100)^{\alpha}} \quad \forall i,j,k \in I \text{ s.t. } i<j<k,\\
    \bar B^{ij}(\alpha,\beta) = \beta  \displaystyle \frac{\sqrt[3]{s_i^2s_j} + \sqrt[3]{s_is_j^2}}{(d_{ij}+100)^{\alpha}} \quad \forall i,j \in I \text{ s.t. } i<j,\\
    \bar B^{i}(\gamma) = \gamma s_i \quad \forall i \in I.
    \end{cases}
\end{equation}
The distance between three different institutions is defined as
$$d_{ijk} = \displaystyle \frac{d_{ij}+d_{jk}+d_{ki}}{3}$$, 
and the normalization factor is given by
\begin{equation}
\bar N(\alpha,\beta,\gamma) = \sum_{i,jk \in I: i<j<k} \bar B^{ijk}(\alpha,\beta,\gamma) + \bar B^{ij}(\alpha,\beta,\gamma) + \sum_{i \in I} \bar B^{i}(\alpha,\beta,\gamma).
\end{equation}
The parameters $\alpha, \beta, \gamma$ are the variables that we constrain to be non-negative real numbers.

\section{Data Availability}\label{Data Availability}
The code used in the current study is available at \url{https://github.com/saraventurini/The-Effects-of-Remote-Working-on-Scientific-Collaboration-and-Impact}.


\section*{Declarations}

\bmhead{Acknowledgements} 
This research was supported in part by Lilly Endowment, Inc., through its support for
the Indiana University Pervasive Technology Institute.
The authors were supported in part by the following grants. S.V.: no. 1927425, no. 1927418, US National Science Foundation (NSF); S.F.: no. 1927425 (NSF), no. 1927418 (NSF) and no. AWD105677 (Novo Nordisk Foundation); S.V., M.M., P. S., and C.R. would also like to thank all members of the MIT Senseable City Consortium (including Atlas University, City
of Laval, City of Rio de Janeiro, Consiglio per la Ricerca in Agricoltura e l’Analisi dell’Economia Agraria, Dubai Future Foundation, FAE Technology, KAIST Center for Advanced Urban Systems, Sondot´ecnica, Toyota, Unipol Tech) for supporting this research.
The authors would like to thank Prof. Vander Freitas for providing the code to assign a single affiliation to authors.

\bmhead{Competing interests}
The authors declare no competing interests.

\bibliographystyle{unsrt}
\bibliography{main-arxiv.bbl} 

\clearpage
\appendix
\renewcommand{\thefigure}{A\arabic{figure}}
\setcounter{figure}{0}  
\section*{Appendix}\label{sec:appendix}

\begin{figure}[H]
\centering
\includegraphics[scale=1.3]{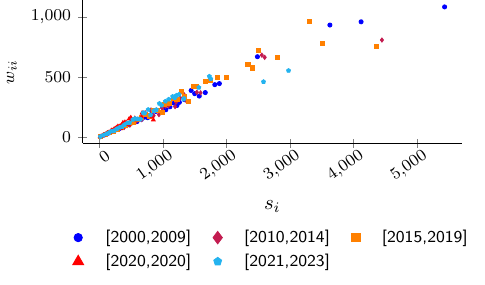}
\caption{
\textbf{Modeling self-edges: institutional strength and collaboration weight}
Relationship between the institutional strength $s_i$ and the weight of self-edges $w_ii$ in the institutional collaboration network, grouped into five consecutive time periods as indicated by the legend. The model assumes that the probability of intra-institution collaboration is proportional to the institution’s strength, which is consistent with the patterns observed in this plot.}
\end{figure}

\begin{figure}[H]
\centering
\includegraphics[scale=1]{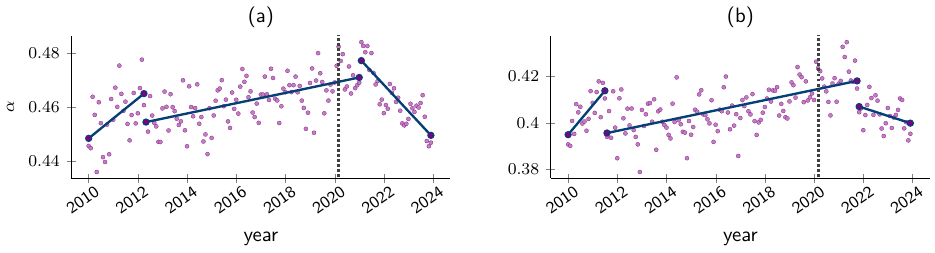}
\caption{\textbf{Temporal evolution of geographical proximity parameter $\alpha$.}
Temporal evolution of the optimized geographical proximity parameter $\alpha$ in the collaboration network.  
\textbf{(a)} Model using overall preprints with pairwise co-authorship connections.  
\textbf{(b)} Model using preprints with three authors, represented as hyperedges. 
A piecewise polynomial model with two breakpoints is fitted by minimizing the squared error.  
The vertical dotted line marks the onset of the COVID-19 pandemic (March 2020).
}
\end{figure}

\begin{figure}[H]
\centering
\includegraphics[scale=1]{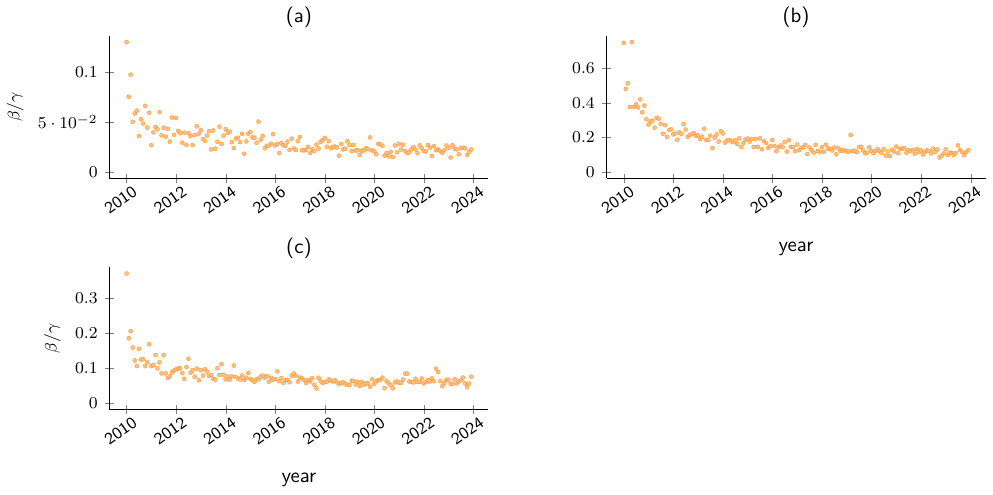}
\caption{
\textbf{Relationship between model parameters $\beta$ and $\gamma$.}
Relationship between $\beta$ and $\gamma$ in the proposed model with two independent variables:
\textbf{(a)} modeling preprints with two authors,
\textbf{(b)} modeling the overall preprints with pairwise
co-authorship connections,
\textbf{(c)} modeling preprints with three authors, represented as hyperedges.}
\end{figure}

\begin{figure}[H]
\centering
\includegraphics[scale=1]{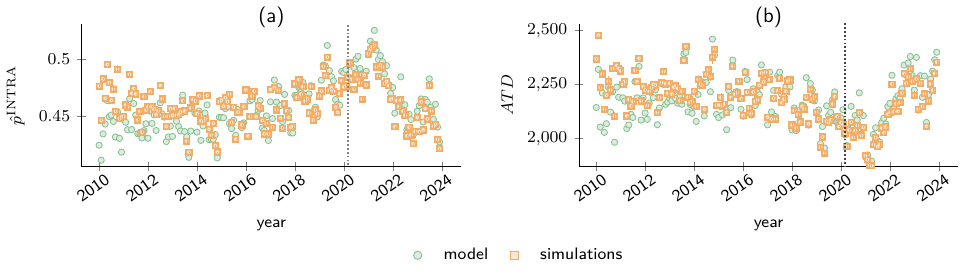}
\caption{
\textbf{Model fit for two-author preprints.}
Modeling preprints with two authors:  
\textbf{(a)} Fitting of the intra-institution collaboration fraction and \textbf{(b)} the average team distance.  
Green indicates the empirical data, while orange represents the mean over 10 simulations, with bars showing the standard error.  
The vertical dotted line marks the onset of the COVID-19 pandemic (March 2020).
}
\end{figure}
\begin{figure}[H]
\centering
\includegraphics[scale=1]{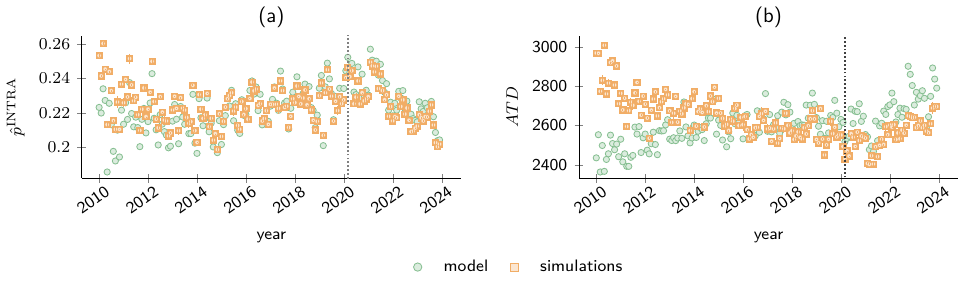}
\caption{
\textbf{Model fit of all preprints.}
Modeling the overall preprints with pairwise co-authorship connections:  
\textbf{(a)} Fitting of the intra-institution collaboration fraction and \textbf{(b)} the average team distance.  
Green indicates the empirical data, while orange represents the mean over 10 simulations, with bars showing the standard error.  
The vertical dotted line marks the onset of the COVID-19 pandemic (March 2020).}
\end{figure}
\begin{figure}[H]
\centering
\includegraphics[scale=1]{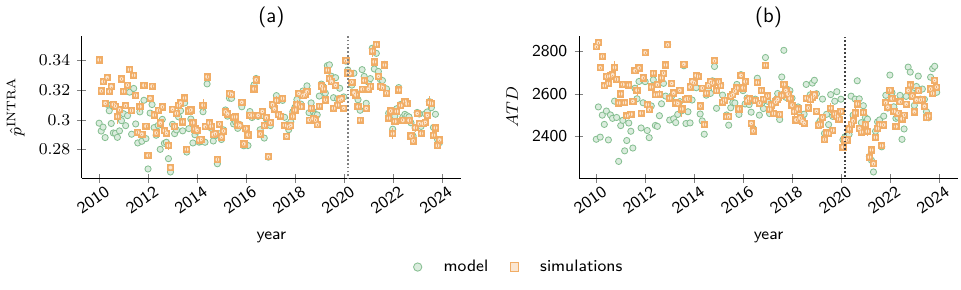}
\caption{
\textbf{Model fit of three-author preprints represented as hyperedges.}
Model using preprints with three authors, represented as hyperedges:  
\textbf{(a)} Fitting of the intra-institution collaboration fraction and \textbf{(b)} the average team distance.  
Green indicates the empirical data, while orange represents the mean over 10 simulations, with bars showing the standard error.  
The vertical dotted line marks the onset of the COVID-19 pandemic (March 2020).}
\end{figure}

\begin{figure}[H]
\centering
\includegraphics[scale=1]{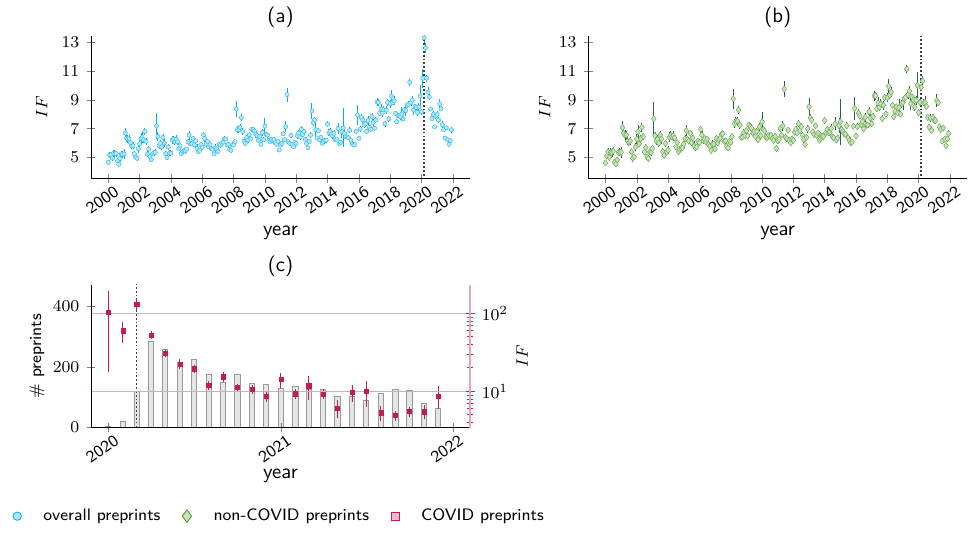}
\caption{
\textbf{Temporal trends in two-year citation counts of preprints.}
\textbf{(a)} Monthly average number of two-years citations received by preprints, with standard error bars (traditional impact factor).  
\textbf{(b)} Same as (a), restricted to non-COVID-related preprints.  
\textbf{(c)} Monthly counts of COVID-related preprints (gray bars) and their average two-years citation counts (magenta, logarithmic scale) with standard error bars. 
Vertical dotted lines indicate the onset of the COVID-19 pandemic (March 2020).}
\end{figure}

\begin{table}[H]
    \centering
    \caption{\textbf{Preprint counts by time period.}  Number of preprints published across five consecutive time periods.}
    \hspace{2cm}
    \begin{tabular}{ccc}
        period & time range & preprints \\
        \midrule
        \RNum{1}&[2000,2009]&284862\\
        \RNum{2}&[2010,2014]&288298\\
        \RNum{3}&[2015,2019]&478754\\
        \RNum{4}&[2020,2020]&139766\\
        \RNum{5}&[2021,2023]&442334\\
    \end{tabular}
\end{table}

\begin{figure}[H]
\centering
\includegraphics[scale=0.88]{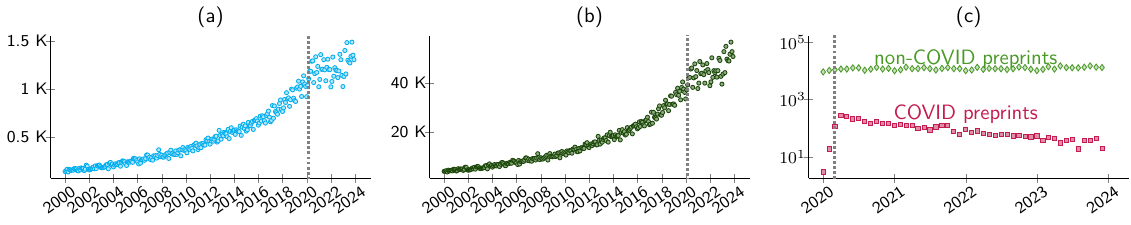}
\caption{\textbf{Temporal trends in preprints and authors.} Time evolution of the number of \textbf{(a)} preprints, \textbf{(b)} authors, and \textbf{(c)} COVID and non-COVID preprints in logarithmic scale.}
\end{figure}

\begin{table}[H]
    \centering
    \caption{ \textbf{Preprints by ArXiv category with counts and percentages.}
    Division of preprints across \texttt{ArXiv} categories, showing both the count and the corresponding percentage ($perc$) for each category. Percentages may exceed $100\%$ in total because preprints can be assigned to multiple categories.}
    \hspace{2cm}
    \begin{tabular}{l|cc}
        category & preprints & perc \\
        \midrule
        Physics &813873 &49.8\%\\
        Computer Science &388934 &23.8\%\\
        Mathematics &338466 &20.7\%\\
        Statistics &76422 &4.7\%\\
        Electrical Engineering and Systems Science &55793 &3.4\%\\
        Biology &28286 &1.7\%\\
        Quantitative Finance &10180 &0.6\%\\
        Economics &4190 &0.3\%\\
    \end{tabular}
\end{table}

\begin{figure}[H]
\centering
\includegraphics[scale=0.88]{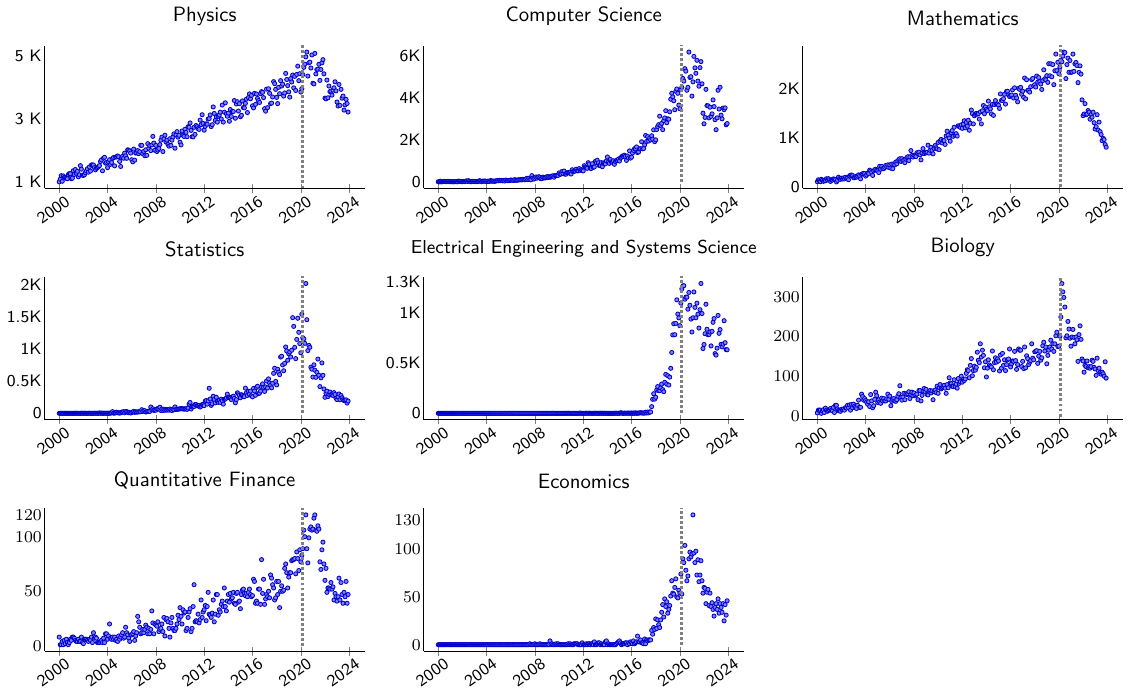}
\caption{\textbf{Temporal trends in preprints by ArXiv category.} Time evolution of the number of preprints into \texttt{ArXiv} categories.}
\end{figure}

\end{document}